\documentclass[pra,twocolumn,amsmath,amssymb,superscriptaddress]{revtex4-1}
\usepackage[english]{babel}
\usepackage{latexsym}
\usepackage{graphics}
\usepackage{epsfig}
\usepackage{color}
\usepackage{bm}
\usepackage{amsmath}
\usepackage{amssymb}
\usepackage{hyperref}

\usepackage{ dsfont }

\usepackage[normalem]{ulem}
\usepackage{color}

\newcommand{\ket}[1]{|#1\rangle}

\newcommand{\bb}[1]{\left( #1 \right)}

\newcommand{\be}{\begin{equation}}
\newcommand{\ee}{\end{equation}}
\newcommand{\bea}{\begin{eqnarray}}
\newcommand{\eea}{\end{eqnarray}}

\usepackage[T1]{fontenc} 

\begin{document}

\title{Roton in a few-body dipolar system}
\author{R.~ O{\l}dziejewski}
\author{W.~G{\'{o}}recki}
\author{K.~ Paw{\l}owski}
\author{K.~Rz\k{a}\.{z}ewski}

\affiliation{Center for Theoretical Physics, Polish Academy of Sciences, Al. Lotnik\'{o}w 32/46, 02-668 Warsaw, Poland}


\begin{abstract}
We solve numerically exactly the many-body 1D model of bosons interacting via short-range and dipolar forces and moving in the box with periodic boundary conditions. We show that the lowest energy states with fixed total momentum can be smoothly transformed from the typical states of collective character to states resembling single particle excitations. In particular, we identify the celebrated roton state. The smooth transition is realized by simultaneous tuning short-range interactions and adjusting a trap geometry. With our methods we study the weakly interacting regime as well as the regime beyond the range of validity of the Bogoliubov approximation.

\end{abstract}


\maketitle

\section{Introduction}
In the 30s of the last century unusual properties of the Helium-II were discovered. 
The subsequent results of Allen and Misener \cite{allen1938}, Kapitza \cite{Kapitza1938} were simulating the development of theoretical models~\cite{london1938,tisza1938, Landau1941, Landau1941b, Landau1947}.
The qualitative theory of superfluidity is due to Landau \cite{Landau1941, Landau1941b, Landau1947}.
He deduced  from the measurement of the specific heat \cite{Kapitza1941} and the second sound velocity \cite{Peshkov1946} that the excitations in the Helium-II must have a peculiar spectrum, with the local minimum \cite{Landau1947}.
The excitation at the local minimum has been called a "roton". Later Feynman alone~\cite{feynman1954} and with Cohen~\cite{Cohen1956} formulated the very first, yet semiquantitative microscopic model explaining the origin of this local minimum.
Finally, in Helium the roton was observed experimentally \cite{henshaw1961}, but rather unsatisfactory agreement between theory and measurement suggested that the exact nature of the rotonic excitation was still missing. It was finally understood many years later by means of subtle ansatzes for the roton's wave function~\cite{Galli1996,Apaja1998}. It should be emphasized that liquid Helium-II is a strongly correlated (with a small condensate fraction) system, where roton's characteristic momentum scales as the interatomic distance. There are still active studies of the roton state in this regime~\cite{Giorgini2013}.

At the beginning of XXI century the roton-maxon spectrum was predicted in completely different physical systems - dipolar gas of ultracold atoms in constrained geometries~\cite{Santos2003,kurizki2003}. Unlike in Helium-II, in this case the interactions are weak and a condensate fraction is dominant. Therefore, one can use the mean field or Bogoliubov description and find the roton state as a Bogoliubov quasi particle~\cite{Santos2003,kurizki2003, fisher2006,ronen2007,pu2008,bohn2009does,parker2009structure,
wilson2010critical,nath2010faraday,klawunn2011local,martin2012,blakie2012roton,
jona2013roton,bisset2013roton,bisset2013fingerprinting,
jona2013time,bohn2013,blakie2013depletion,natu2014dynamics}. The dispersion curve of such systems is related to a specific k-dependence of an effective interaction potential rather than to strong correlations. Possibility of changing the particles polarization as well as almost free tuning of the short range interactions combined with the trap geometry modifications enables unprecedented flexibility in the study of the roton spectrum in dipolar gases~\cite{Santos2003,kurizki2003, fisher2006,ronen2007,pu2008,bohn2009does,parker2009structure,
wilson2010critical,nath2010faraday,klawunn2011local,martin2012,blakie2012roton,
jona2013roton,bisset2013roton,bisset2013fingerprinting,
jona2013time,bohn2013,blakie2013depletion,natu2014dynamics} ending with a recent experimental confirmation of the phenomenon~\cite{chomaz2018}. Usually the dipolar system is studied within the Bogoliubov approximation, so that there is no access to the detailed structure of the low lying excitations. Only a few many-body investigations were performed for the roton state using different techniques~\cite{boronat2007,boronat2009,hufnagl2009, pupillo2010}. A good attempt can be made by a numerically exact solution to a many-body problem with a rotonic characteristic. Even if found for relatively small number of particles, modern experiments with a precise control over only a few atoms in optical lattices or single traps (see for instance~\cite{greiner2002quantum,serwane2011deterministic,
meinert2015probing,baier2016extended,baier2018realization}) allow to test its physical predictions.

In this work we present numerically exact results for a quasi-1D model, which admits the roton excitation. In a number of recent papers low excitation of 1D interacting bosons were already investigated (see for instance~\cite{Astrakharchik2005,caux2006dynamical,Astrakharchik2008,
de2008low,imambekov2012one,Astrakharchik2012} and references therein). The historically earliest example is the famous Lieb-Liniger model \cite{LiebLiniger1963, Lieb1963} comprising of $N$ contact interacting bosons moving on the circle. Their seminal analytical solution predicts two branches of elementary excitations, which was also observed experimentally~\cite{meinert2015probing}. The upper type-I excitation branch was immediately recognized as the Bogoliubov excitation spectrum \cite{fetter2012}.  The states of the lower type-II excitation branch, were identified later with grey and dark solitons arising in the mean field theory of ultracold gases \cite{KanamotoCarr2008, KanamotoCarr2010,syrwid2015,syrwid2016}. Little is known about the classification of exact many-body elementary excitations in the dipolar gas. For (quasi)-1D model with bosons interacting only by repulsive dipolar interactions the lowest energy states resemble rather type II excitations from the Lieb-Liniger model~\cite{de2008low,oldziej2018} and for at least weakly interacting regime the picture with two branches of elementary excitations is also expected in this case~\cite{oldziejewski2018many}. On the other hand, in the dipolar systems, well understood Bogoliubov spectrum may exhibit a local minimum identified as a roton~\cite{Santos2003,kurizki2003}. When the interatomic forces are of the attractive character on the short-scale, whereas the long-range part of potential is repulsive, the interplay of these two interactions may lower the energy of the roton mode even to the ground state level. It opens a significant question: is it possible in a dipolar analogue of the Lieb-Liniger model that the two branches cross, such that it is a type-I Bogoliubov excitation, in particular the roton, which would appear in the lower branch?

It is a purpose of this work to show that by tuning short-range interactions and adjusting a ring geometry one can continuously change~\cite{fialko2012nucleation} the character of the lowest energy state for a given total momentum of the system from a type-II excitation to the roton mode. We also analyze a numerically exact roton's wave function in the weakly interacting regime and its position and momentum properties.


\section{Model}
We consider $N$ dipolar bosons confined in both transverse directions $\hat{y}$ and $\hat{z}$ 
with a tight harmonic trap of a frequency $\omega_\perp$. 
Multi-particle wave-function is approximately the Gaussian in tight directions for all variables.
It requires the chemical potential $\mu$ much smaller than energy of the first excited state in the transverse direction, $\mu \ll \hbar \omega_\perp$.  In the longitudinal direction $\hat{x}$ the space is assumed to be finite, with the length $L$ and with the periodic boundary conditions imposing quantisation of momenta in that direction. All atoms are polarized along the $\hat{z}$ axis. The above system corresponds to atoms moving on the circumference of a circle, having the dipole moments perpendicular to the circle-plane. Hence, in analogy with nuclear physics \citep{Mottelson1999,HAMAMOTO199065} and following \cite{KanamotoCarr2008,KanamotoCarr2010} we call the lowest energy states of a given total momentum of the system, the yrast states. Our quasi-1D system is governed by Hamiltonian
\begin{equation}\label{Ham}
\hat{H} = \sum_{k}\frac{\hbar^2k^2}{2m}\hat{a}_{ k}^\dagger \hat{a}_{k}\\
+\frac{1}{2L}\sum_{k_1,k_2, k} \hat{a}_{ k_1+k}^\dagger \hat{a}_{k_2-k}^\dagger V_{{\rm eff}}(k) \hat{a}_{k_1}\hat{a}_{k_2},
\end{equation}
 with  $\hat{a}_k$ ( $\hat{a}_k^{\dagger}$) anihilating (creating) a boson with momentum $k$. The effective potential consists of the long-range dipolar part and the short-range part, namely $V_{{\rm eff}}(k) =V_{{\rm sr}}(k)+V_{{\rm dd}}(k)$.
 
 The quasi-1D dipolar potential reads $V_{{\rm dd}}(k)=\frac{3\hbar^2 a_{\rm dd}}{m l_{\perp}^2} \bb{ 1 +  f\bb{( l_{\perp}k)^2 /2}}$ with $l_{\perp}=\sqrt{\hbar/m\omega_{\perp}}$. The parameter $a_{\rm dd} = m\mu_0 d^2/(12 \pi\hbar^2)$ is a "dipole length", where  $d$ is an atomic dipole moment and $\mu_0$ is the vacuum permeability. 
This effective quasi-1D potential comes from integration of the full $3D$ dipolar interaction over both  transverse variables.
The singular part coming from this integration is incorporated with the short range interaction. The function $f$ which appears in Eq. \ref{Ham} is equal to $f(u) = u\,e^{u}{\rm Ei}(u)$, where Ei is the exponential integral \cite{abramowitz}.\\
\indent Stability of our calculations requires smoothing of a usual short range interaction model used in the ultracold physics, the delta function.  We choose a Gaussian model~\cite{doganov2013two,von2008energetics,blume2012few,
christensson2009effective,klaiman2014breaking,imran2015exact,
beinke2015many,bolsinger2017beyond,bolsinger2017ultracold}, namely $V_{{\rm sr}}(k)= V_0 e^{-\frac{1}{2} k^2 r^2}$ with $r$ standing for the  potential range and $ |V_0|$ for its depth. This step makes our model more realistic, imitating the attractive van der Waals interaction. For convenience we set $V_{0}=\frac{\hbar^2 a}{m l_{\perp}^2}$ with $a$ mimicking an usual scattering length. The relation between Gaussian model and the real scattering length can be found in~\cite{jeszenszki2018s} and references therein. Below we use box units where $L/2\pi$, $2\pi\hbar/L$ and $4\pi^2\hbar^2/mL^2$ are the units of length, momentum and energy respectively.



Our effective potential $V_{\rm{eff}}(k)$ corresponds to calculating the interactions along the circumference positions with the periodicity of the system accounted for. However, we checked that it would not be changed significantly if a bit elegant but more realistic geometric distance over the chord was used (see Appendix \ref{Sec:rvsp}).

We access the many-body eigenstates of Hamiltonian \eqref{Ham} by exact diagonalization using the Lanczos algorithm~\cite{lanczos1950iteration}. Our calculations are performed in the Fock space spanned by the plane-wave basis with a maximum total kinetic energy of the system $E_{\rm max}=k_{\rm max}^2/2$ -- with  single-particle momentum $k_{\rm max}\gg 1/r$ -- sufficiently high to assure convergence. Here we employ the fact that the total momentum of the system $\hat{K}=\sum_k \,k\, \hat{a}_{k}^\dagger \hat{a}_{k}$ is conserved, $\left[ \hat{H}, \hat{K} \right]=0$, so its eigenvalues $K$ are  good quantum numbers, used here together with the total number of atoms $N$ to label different eigenstates $\ket{N,\, K,\,i}$ enumerated by $i$ with $i=0$ corresponding to an yrast state. We remind the Reader that for finite systems on the ring it suffices to consider the eigenstates only up to $ K/N =1/2$~\cite{Lieb1963,KanamotoCarr2010}. This comes from the presence of the so called {\textit{umklapp}} process~\cite{Lieb1963}. Any eigenstate with a total momentum $K'=p\cdot N+K$ (where $p\in\mathds{Z}, -\frac{N}{2}\leq K\leq\frac{N}{2}$) may be understood as the state with a total momentum $K$ with a shifted center-of-mass momentum (see Fig. 5 in~\cite{KanamotoCarr2010}). Note that such shifting does not change the internal structure of the state.

\section{Results}
\begin{center}
\begin{figure}[htb!]
	\includegraphics[width=0.36\textwidth]{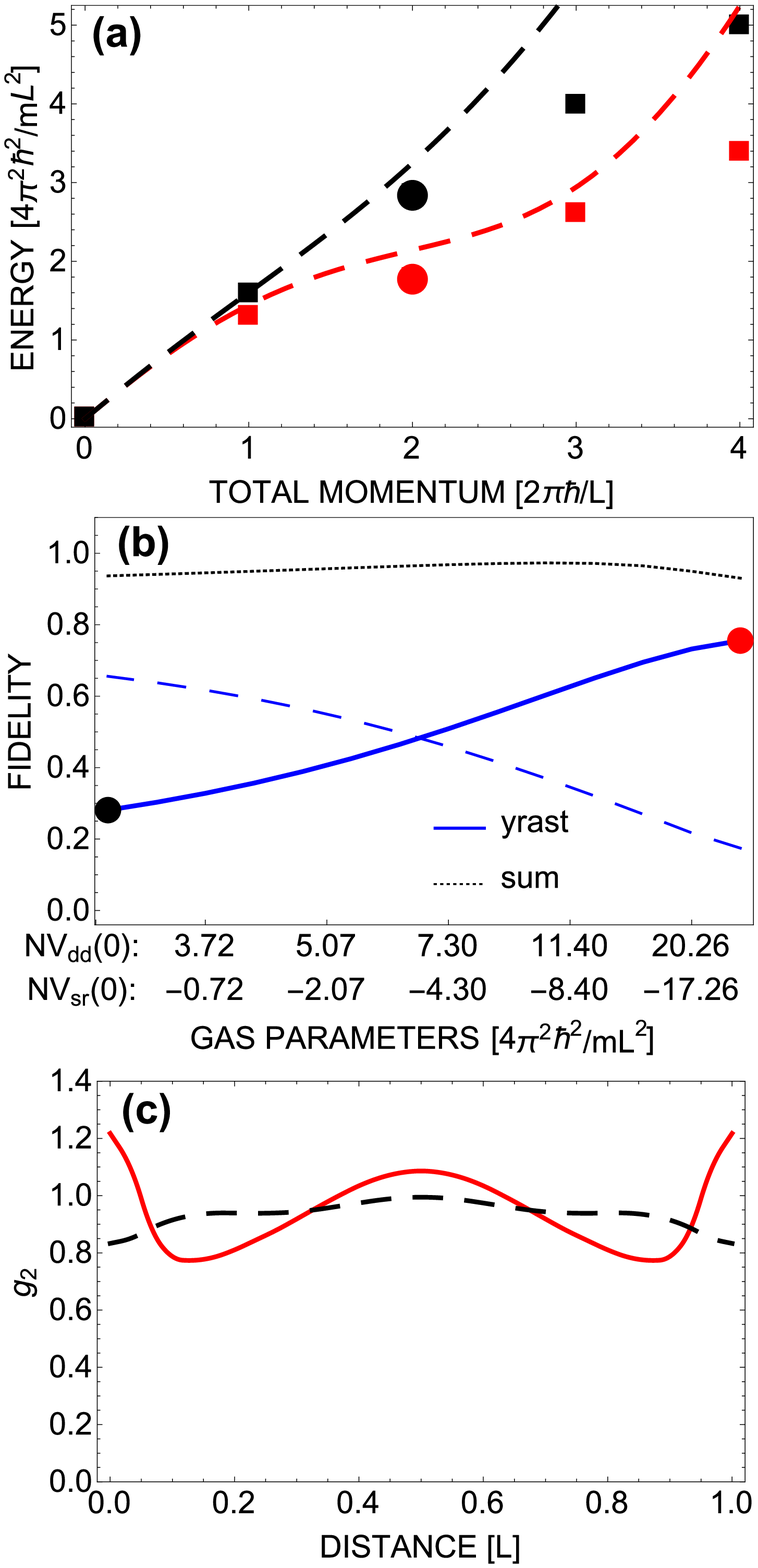} \\
	\includegraphics[width=0.44\textwidth]{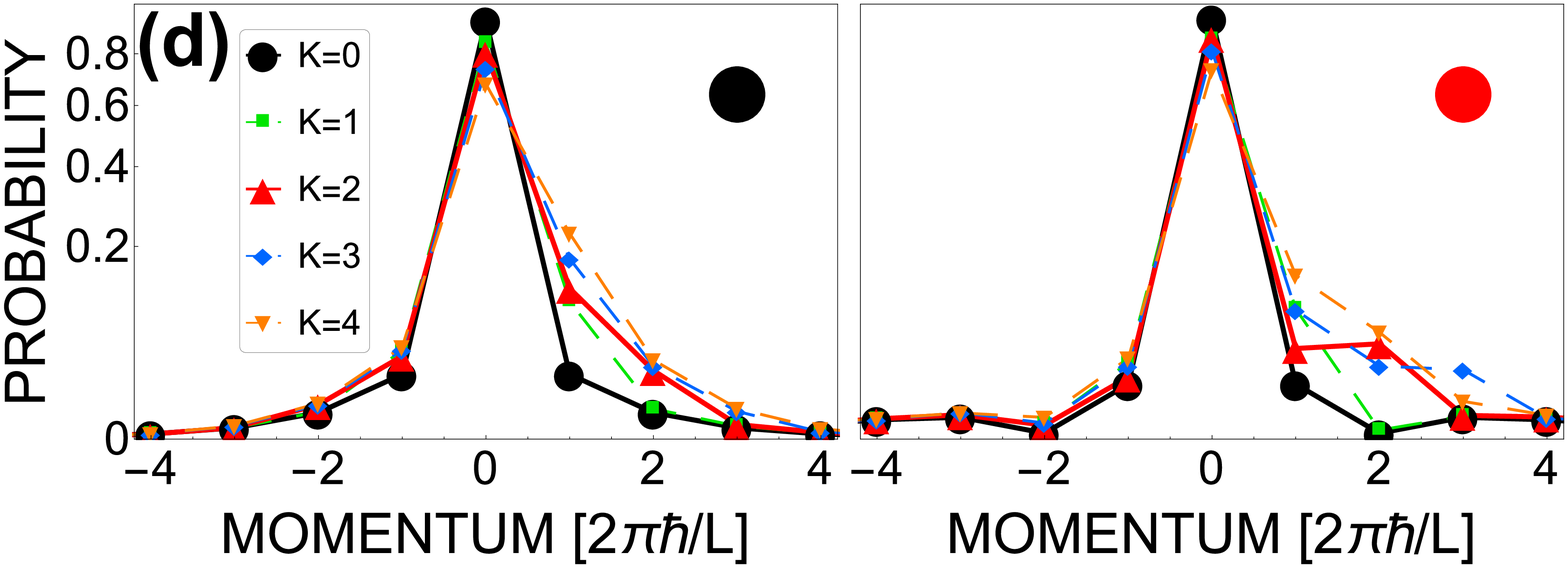}
	
	\caption{(color online) Results for weak interactions (a) Energy of the yrast states for $a=0$ and $\omega_{\perp}\approx 2\pi \times 41$ kHz (black squares) and $a\approx -378\, a_{0}$ and $\omega_{\perp}\approx 2\pi \times 365$ kHz  (red squares)  for $N=16$ dysprosium atoms ($a_{\rm dd}=132$ $a_{0}$ and $r= 182$ $a_{0}$) as a function of the total momentum compared with the corresponding Bogoliubov excitation spectrum (dashed lines). (b) Fidelities between the first two eigenstates and Bogoliubov excitation for $K=2$ as a function of $a$ and $\omega_{\perp}$ ($N$ and $a_{\rm dd}$ are constant and as in (a)). (c) The normalized second order correlation function as a function of a distance for two states from (b) marked by color filled circles. (d) Single-particle momentum probability $P(k)$ for all states from (a) (five for each spectrum).
	 \label{fig:Fig1}}
\end{figure}
\end{center}

\begin{center}
\begin{figure}[htb!]
	\includegraphics[width=0.36\textwidth]{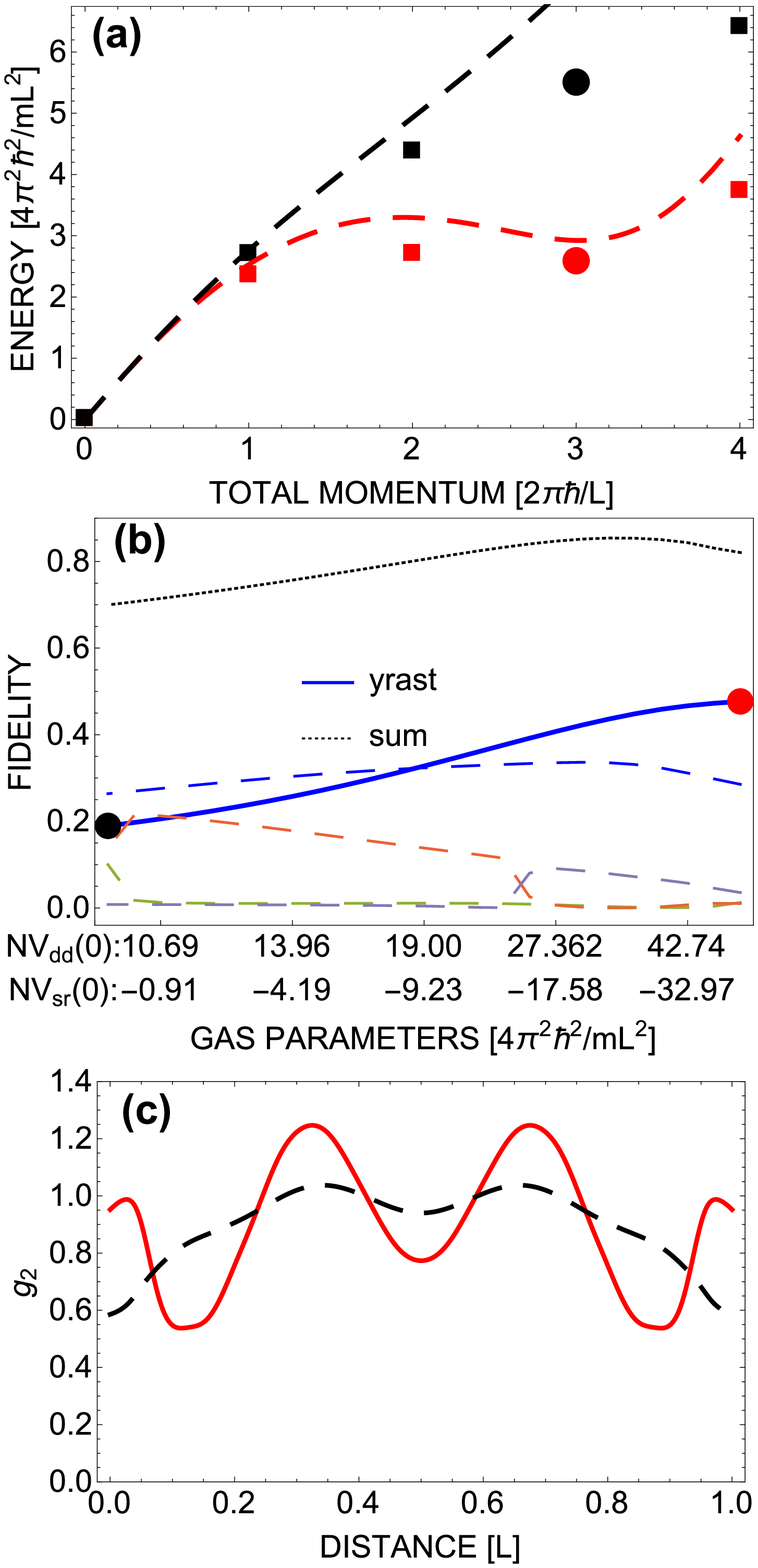} \\
	\includegraphics[width=0.44\textwidth]{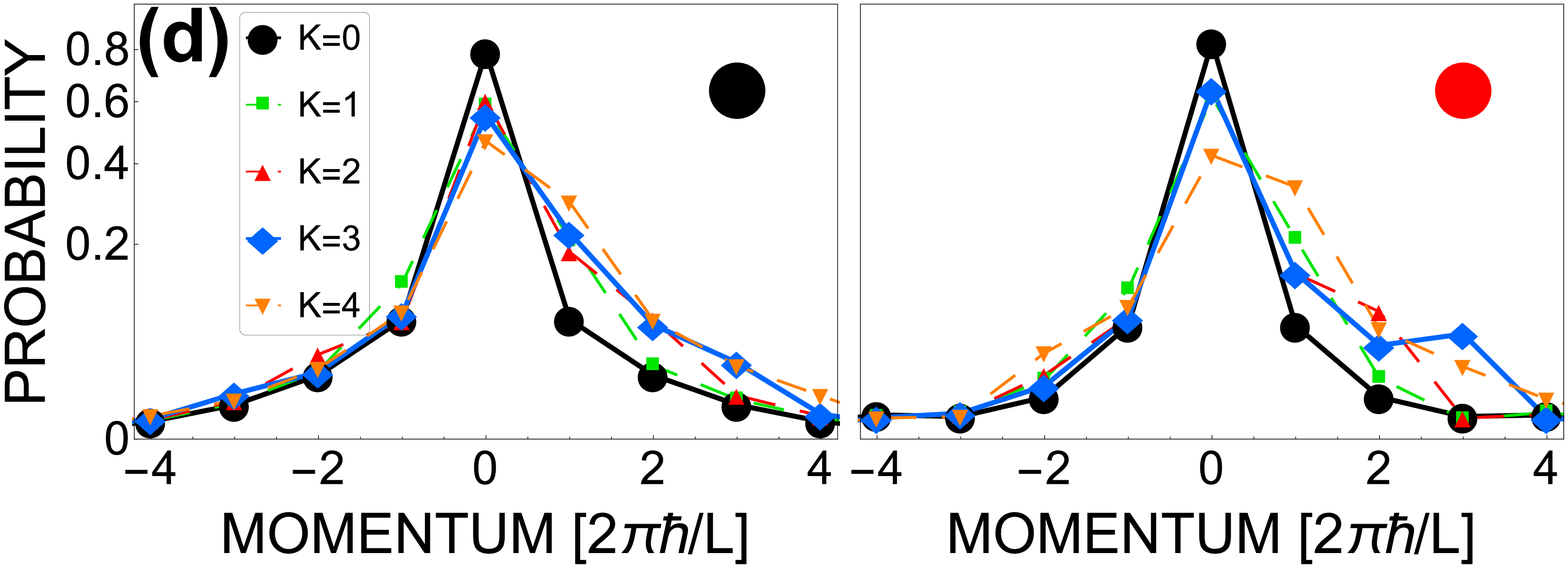}
	\caption{(color online) Results for weak interactions (a) Energy of the yrast states for $a=0$ and $\omega_{\perp}\approx 2\pi \times 35$ kHz  (black squares) and $a\approx -2080\, a_{0}$ and $\omega_{\perp}\approx 2\pi \times 190$ kHz (red squares)  for $N=10$ atoms ($a_{\rm dd}=792$ $a_{0}$ and $r= 272$ $a_{0}$) as a function of the total momentum compared with the corresponding Bogoliubov excitation spectrum (dashed lines). (b) Fidelities between the first five eigenstates and Bogoliubov excitation for $K=3$ as a function of $a$ and $\omega_{\perp}$ ($N$ and $a_{\rm dd}$ are constant and as in (a)). (c) The normalized second order correlation function as a function of a distance for two states from (b) marked by color filled circles. (d) Single-particle momentum probability $P(k)$ for all states from (a) (five for each spectrum).
	 \label{fig:Fig2}}
\end{figure}
\end{center}

In the following paragraphs of this work we consider two different situations, namely with weak interactions where the depletion (given by $P(K=0)$ in Fig.~\ref{fig:Fig1}d and \ref{fig:Fig2}d) of a ground state is less than 5\% and stronger interactions where its value is around 20\%. Note, that the latter case is still far from the Helium-II regime. We present in Fig.~\ref{fig:Fig1} our analysis of yrast states for the first situation. We consider $N=16$ dysprosium atoms with $a_{\rm dd}=132$ $a_{0}$ and the potential range $r= 182$ $a_{0}$, where $a_0$ is the Bohr radius. We initially set $a$ and $\omega_{\perp}$ corresponding to the usual situation where the yrast states energies clearly do not follow the Bogoliubov spectrum (black dashed line) given by: 
\begin{equation}
\epsilon_k=\sqrt{\frac{k^2}{2}\bb{\frac{k^2}{2}+2NV_{\rm eff}(k)}}
\end{equation}
and rather resemble the lowest excitation branch from the Lieb-Liniger model~\cite{oldziejewski2018many,oldziej2018}(black squares in Fig.~\ref{fig:Fig1}a). Then we continuously change $a$ and $\omega_{\perp}$ (a simillar effect would be observed if one changed the length of the box) keeping $V_{\rm{eff}}(0)={\rm {const.}}$ We finally end with the profoundly different spectrum (red points in Fig.~\ref{fig:Fig1}a) closer to a corresponding Bogoliubov dispersion relation (red dashed line), in particular with the characteristic inflection point for $K=2$. Our result suggest that at least some of yrast states may change their character from collective type-II excitations~\cite{oldziej2018} to type-I ones. Moreover, we argue that the inflection point can be identified with the roton-like state.

To test our hypothesis about the change of the character of the yrast state for $K=2$ we compare it together with the first excited state with the number conserving Bogoliubov approximation \cite{dum1998} sketched here briefly. The spectrum in the Bogoliubov approximation is explained by the concept of quasiparticles that, in our case, has to be rewritten in terms of Fock states in particle basis. We use the following Ansatz \cite{Leggett2001} for the
Bogoliubov vacuum ($K=0$):
\begin{equation}\label{vac}
\ket{0}_{B} \propto \left( \left( \hat{a}_0^{\dagger} \right)^2 - 2 \sum\limits_{k>0}^{\infty}\frac{v_{k}}{u_{k}} \hat{a}_k^{\dagger}\hat{a}_{-k}^{\dagger}\right)^{N/2}\ket{\rm vac},
\end{equation}
where $\ket{\rm vac}$ is the particle vacuum and $u_{k},v_{k}=\left( \sqrt{\epsilon_k/E_k}\pm \sqrt{E_k/\epsilon_k}\right)/2$ with $E_k=k^2/2$. A single Bogoliubov excitation with a total momentum $K$ is expressed by $\ket{N,K}_{B}\propto \left( u_K \hat{a}_{0}\hat{a}_{K}^{\dagger}+ v_K \hat{a}_{0}^{\dagger}\hat{a}_{-K}  \right)\ket{0}_B$. To trace a continuous transformation of the yrast state from type-II to type-I excitation we evaluate the fidelities $\left| \langle {N,K,i} \ket{N,K}_B \right|^2$, which is depicted in Fig.~\ref{fig:Fig1}b. For the initial values of the parameters $a$ and $\omega_{\perp}$ the first excited state is a Bogoliubov excitation, whereas the yrast state remains a type-II excitation~\cite{oldziej2018}- a fact observed in the Lieb-Liniger result as well. Then we observe a gradual exchange of the states' character as we modify the effective potential ending with a complete role reversal of the two first states. Note, that the sum of the fidelities (black dotted line in Fig.~\ref{fig:Fig1} b) is almost equal to 1 at any stage of the transition. It means that Bogoliubov excitation, to a good approximation, remains in a plane spanned by the two lowest eigenstates.

To show the qualitative change of the yrast state for $K=2$ we calculate the normalized second order correlation function 
$g_2(z) := \langle \Psi^{\dagger}(z)\Psi^{\dagger}(0)\Psi^{}(0)\Psi^{}(z) \rangle /\langle \Psi^{\dagger}(z)\Psi^{}(z) \rangle\langle \Psi^{\dagger}(0)\Psi^{}(0) \rangle $ (Fig.~\ref{fig:Fig1}c), which can be measured in experiments with ultracold atoms, see for instance \cite{bloch2005,westbrook2005,bouchoule2006,westbrook2007}. We observe a dramatic difference between two yrast states for border cases from Fig.~\ref{fig:Fig1}b (marked as black and red points). Namely that almost flat distribution typical for type-II excitation in weakly interacting regime is replaced by a function exhibiting an enhanced regular modulation with the number of maxima given by $K_{\rm rot}$, which is the roton momentum.

Note, that for a small number of particles we are able to find stable solutions corresponding to realistic, physical gas parameters ($NV_{\rm{sr}}(0)= -23.98$, $N V_{\rm dd}(0)= 26.98$) only for the Bogoliubov spectrum with the inflection, not to the one with the characteristic local minimum. Using the gas parameters for which the Bogoliubov spectrum has the local minimum implies much stronger interactions for our few-body system.
Our result would approach Bogoliubov's predictions in the limit of $N\rightarrow \infty$ (see Appendix \ref{Sec:infty}).

Instead of going to larger systems, we turn to the strong interactions scenario with $N=10$, which is beyond the Bogoliubov approximation. In Fig. \ref{fig:Fig2} we summarize our findings, where the characteristic local minimum for $K=3$ is present. In this case the spectrum is calculated with an accuracy of several percent only. Our previous conclusions hold also for this situation. However, the role reversal of the lowest states is more subtle because of higher momentum of the roton. At the end of the transition we stay with the yrast state, which still has the overlap with Bogoliubov (50\%) and at the same time exhibits the enhanced regular modulation in second order correlation function and the local minimum in the spectrum. It is the roton-like state in a regime between weak interaction and Helium-II regime.

Note that the presence, position, and depth of
the roton minimum for both cases considered in this work are tunable by varying the number of atoms $N$,
trapping frequency $\omega_{\perp}$ and the short-range coupling
strength as it was predicted for the roton state in the meanfield studies of ultracold dipolar gases~\cite{Santos2003}. We choose $K_{\rm rot}/N<1/2$ to minimize the impact of the \textit{umklapp} process ~\cite{Lieb1963}, discussed earlier in this work, on the eigenstates.

To fully comprehend the difference between the two types of low-energy excitations we study the probability $P(k)=\frac{1}{N}\langle\hat{a}^{\dagger}_k\hat{a}_k\rangle$ of finding a single-particle moving with momentum $k$ for yrast states with various $K$. For both weak and strong interactions the type-II yrast states (black markers in Fig. 1d and 2d) for $K>1$ beside $k=0$ mainly consists of $k=1$ states, which is more visible as we increase $K$. It corresponds to a dominant role of one of the Dicke states (exactly $K$ atoms with $k=1$ and $N-K$ with $k=0$) in their many-body wave function~\cite{oldziejewski2018many,oldziej2018}, especially for weak interactions. On the other hand, in the rotonic cases (red markers in Fig. 1d and 2d) we observe a local maximum of $P(k)$ for $k=K_{\rm rot}$ for the yrast states with $K_{\rm rot}$, which clearly resembles recently published result by F. Ferlaino's group \cite{chomaz2018}. It means that the yrast state for $K_{\rm rot}$ has a single particle excitation character rather than a collective one, so that within our, experimentally achievable, procedure one can completely change the character of the low-energy excitations.
\section{Discussion}
We find with our numerically exact treatment that all the properties of the roton state discussed earlier can be understood by analysing contributions of different Fock states to its wave function. In both cases of interactions studied in this work, we find that the dominant contribution to the roton states comes from the so called $W$ state
$\ket{0_{-k_{max}},...(N-1)_{0},0_{1},1_{2},...,0_{k_{max}}}$, as one would expect for the Bogoliubov excitations~\cite{oldziejewski2018many}. The latter state is important from the fundamental point of view, as representative of an entanglement class \cite{Wstate2000}, and applied side -  it can be used to beat the standard quantum limit for the metrological tasks \cite{Pezze2009}. The state was recently produced via non-demolition measurement \cite{Haas2014}. According to our earlier findings~\cite{oldziejewski2018many}, which holds also for purely dipolar repulsion~\cite{oldziej2018}, the low-lying excitations of weakly interacting bosons are highly-entangled states dominated by the Dicke state, a result of the bosonic statistics mainly. However, the interplay between the short-range and long-range interactions of the opposite sign can promote the excitation with the dominant $W$ state as a low-lying excitation for $K>1$ in the system.
\section{Conclusion}
To summarize, we showed that manipulating physical parameters in our model one can continuously alter the character of a given yrast state from type-II excitation to the roton mode. We emphasise the fact that the effect is already present in relatively small systems enabling use of the simplest exact diagonalization of the whole Hamiltonian. All interesting properties of the roton-like mode both in the momentum and the position representations come from the fact, that the $W$ state plays the dominant role in the roton state in the plane wave basis. It is in the stark contrast to the weakly repulsive bosons, where the dominant role of the Dicke states is observed~\cite{oldziejewski2018many,oldziej2018}. We show that the normalized second order correlation function, accessible in experiments, displays characteristic enhanced regular modulation for the roton state. Within our many-body model we access stronger regimes between the weakly interacting one and the Helium-II scenario, finding the roton mode also in this case. Our results open new questions concerning quasi-1D systems with both long-range and short-range interactions. Is it possible to fully replace type-II branch with type-I branch as low-lying excitations? Would solitonic branch still exist in the spectrum? The thermodynamic properties of dipolar bosons were investigated only approximately, in the weakly interacting regime~\cite{Bisset2011,Bisset2012,Ticknor2012}. The results presented in this paper can motivate further research in this direction, but using full many-body approach accounting for the lower branch and the transitions discussed here.


\begin{acknowledgements}
We acknowledge fruitful discussions with K. Sacha, A. Syrwid, A. Sinatra and Y. Castin. This work was supported by the (Polish) National Science Center Grants 2016/21/N/ST2/03432 (R.O. and W.G.), 2014/13/D/ST2/01883 (K.P.)  and 2015/19/B/ST2/02820 (K.R.).
\end{acknowledgements}

\appendix

\section{The effective potential. Realistic vs. periodic \label{Sec:rvsp}}

\begin{center}
\begin{figure}[htb!]
	\includegraphics[width=0.44\textwidth]{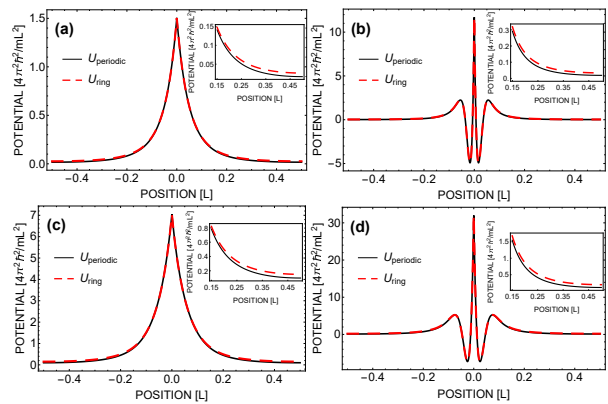} 
	
	\caption{(color online) Comparison between the effective potential $U_{\rm periodic}(x)$ calculated with periodicity accounted for (black line) and $U_{\rm ring}(x)$ with the distance over the chord (red dashed line). (a) Parameters as for the black squares from Fig.~\ref{fig:Fig1}a. (b) Parameters as for the red squares from Fig.~\ref{fig:Fig1}a. (c) Parameters as for the black squares from Fig.~\ref{fig:Fig2}a. (d) Parameters as for the black squares from Fig.~\ref{fig:Fig2} a. Insets: Magnification of the region, where the difference between two methods are the most significant.
	 \label{fig:pote}}
\end{figure}
\end{center}

In the main text we use the effective potential (in the momentum representation) $V_{\rm{eff}}(k)$, that originates as follow. For quasi-1D model on the infinite line, the effective potential in the space representation  $U_{\rm 1D}(x)$ takes the form:
\begin{equation}
\begin{split}
&U_{\rm1D}(x)=\frac{\hbar^2a}{ml_\perp^2}\frac{g(x/r)}{r}+\frac{\hbar^2a_{dd}}{ml_\perp^2}\frac{h(x/l_\perp)}{l_\perp}\quad {\rm where}\\
&g(q)=\frac{1}{\sqrt{2\pi}}e^{-\frac{x^2}{2r^2}},\\
&h(q)=\frac{3}{4}\Big(-2|q|+\sqrt{2\pi}(1+q^2)e^{\frac{q^2}{2}}{\rm{Erfc}}\Big(\frac{|q|}{\sqrt{2}}\Big)\Big).
\end{split}
\end{equation}
As we consider a finite system with the periodic boundary conditions, we introduce $U_{\rm periodic}(x)=\sum_{n\in\mathds{Z}}U_{\rm1D}(x+nL)$. From \textit{Poisson summation formula} it satisfies $U_{\rm periodic}(x)=\frac{1}{L}\sum_{k\in\frac{2\pi}{L}\mathds{Z}}e^{i k x}V_{\rm eff}(k)$
 (where $V_{\rm eff}(k)=\mathcal{F}(U_{\rm 1D})(k)$). However, if one wants to deal with a real ring with atoms moving on its circumference the effective potential should rather depend on the geometric distance over the chord $U_{\rm ring}(x)=U_{\rm1D}\Big(\frac{L}{\pi}\sin(\frac{\pi x}{L})\Big)$. In Fig.~\ref{fig:pote} we compare both approaches. As we see both curves are almost indistinguishable in the regions where the value of the effective potential is meaningful. A very small difference in all cases from Fig.~\ref{fig:pote} is observed only on the potential tail.

\section{Convergence towards $N\rightarrow \infty$ limit \label{Sec:infty}} 

\begin{center}
\begin{figure}[htb!]
	\includegraphics[width=0.44\textwidth]{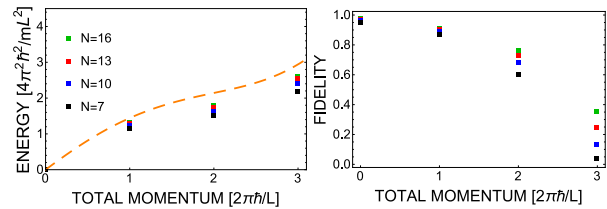} 
	
\caption{(color online) Upper panel: Energy of the yrast states as a function of the total momentum compared with the corresponding Bogoliubov excitation spectrum (orange dashed line) for different number of atoms, from top to bottom N = 16 (green), 13 (red), 10 (blue), 7 (black). 
Gas parameters for all the results are the following (in the box units defined in the main text): $NV_{sr}(0)= -23.98$, $NV_{dd}(0)= 26.98$. For $N=16$ it corresponds to parameters from Fig. \ref{fig:Fig1} (red squares spectrum). Bottom panel: Fidelities between the yrast states and Bogoliubov excitations for the yrast states from the upper panel. Color coding and parameters as in the upper panel (from top to bottom: N = 16, 13, 10, 7).
	 \label{fig:enafid}}
\end{figure}
\end{center}

In the Bogoliubov approximation one operates with the gas parameters $NV_{\rm{sr}}(0)$, $N V_{\rm dd}(0)$ (in the box units defined in the main text) with the assumption of weak interactions and large number of atoms $N$.
Obviously, in the many-body approach, where $N$ is finite, the energy of the pairwise interactions is significantly higher. Then, one can ask how many atoms (how weak interactions) one needs to converge with the many-body solution towards $N\rightarrow \infty$ limit. To answer it, we study energies of a series of yrast states (left panel of Fig.~\ref{fig:enafid}) and their overlaps with the corresponding Bogoliubov excitations given by fidelities (right panel of Fig.~\ref{fig:enafid}) defined in the main text. We obtain both the spectrum and the fidelities for different number of atoms $N$ ranging from 7 to 16. The parameters for different $N$ are chosen to always produce the same Bogoliubov excitation spectrum with the inflection point as for red dashed line in Fig. \ref{fig:Fig1}a in the main text. We see that even for small number of atoms $N=16$ we obtain very good overlap with the Bogoliubov approximation, especially for $K\leq 2$. However, our numerically exact solution includes all the possible correlations between atoms, hence it cannot be fully reproduced by single Bogoliubov excitation.

\bibliographystyle{apsrev4-1}
\bibliography{bibliography}

\end{document}